\documentstyle[preprint,aps]{revtex}
\begin{document}
\draft
\preprint{IU-MSTP/22, hep-th/9707161}
\title{%
Resummation of Vacuum Bubble Diagrams\\
in Gaussian Propagator Model}
\author{Hirofumi Yasuta\cite{email}}
\address{%
Department of Physics, Ibaraki University, Mito 310, Japan}
\date{\today}
\maketitle
\begin{abstract}
An imaginary part of the false-vacuum energy density in a metastable
system, i.e., the decay width due to quantum tunneling, might be
reproduced by Borel resummation of vacuum bubble diagrams. We examine
the convergence of this prescription in the Gaussian
propagator model, in which the analytical expression of vacuum
bubbles to the ninth order of loop expansion is available.
\end{abstract}
Usually, one uses the bounce calculus~\cite{inst} to estimate
the decay width of a metastable system due to quantum tunneling.
This is consistent with the conventional picture that the quantum
tunneling is a purely non-perturbative phenomenon. However, several
methods have been proposed so far~\cite{K,SY,T} which might allow us
to handle tunneling effects with using a conventional perturbation
series. According to these approaches, a certain class of physical
systems seems to have in the perturbation series the whole physical
information, including higher order corrections to the leading bounce
approximation.

In~\cite{SY}, Suzuki and the present author proposed such a
perturbative method to reproduce an imaginary part of the
false-vacuum energy density. The method is based on Borel
resummation~\cite{Z} of vacuum bubble diagrams and is applicable to
super-renormalizable field theoretical models. The validity of the
method has been tested in one-dimensional (i.e., quantum mechanical)
and two-dimensional $O(N)$~symmetric $\lambda\phi^4$~model~\cite{SY}.
In the former, the imaginary part of the ground state energy was
successfully reproduced in a high accuracy. In the latter,
unfortunately, a shortage of the perturbation series prevents us to
draw a quantitative conclusion on the convergence of the method.

In this note, we test the convergence of our proposal by applying it
to the scalar $\phi^4$ theory with a Gaussian propagator~\cite{BDZ}.
The Euclidean action is given by
\begin{equation}
   S[\phi]=\int d^Dx\,\biggl[
   \frac{1}{2}\phi(x)e^{-\Delta}\phi(x)
   -\frac{g}{4!}\phi^4(x)\biggr],
\label{one}
\end{equation}
where $\Delta$~is $D$-dimensional Laplacian. This model has
advantages for our purpose: Since the propagator in momentum space is
\begin{equation}
	G(p)=e^{-p^2},
\label{two}
\end{equation}
an evaluation of Feynman diagram is just Gaussian integrations and no
UV divergences present. These facts allow the evaluation of Feynman
diagrams to a relatively high order in an arbitrary dimensional
space~\cite{BDZ}. This model also allows to carry out the analytical
bounce calculus~\cite{BDZ}. Unfortunately, the kinematical meaning of
this model is not clear due to the odd-looking ``kinetic term,''
(actually, the classical equation of motion has an infinite order
time derivative) and thus the meaning of ``tunneling'' or ``decay''
is not obvious. Nevertheless, we adopt this model as the testing
ground to investigate the convergence of our prescription for Borel
resummation of vacuum bubble diagrams.

We shall consider a ``metastable case'' ($g>0$) and study a sum of
vacuum bubble diagrams:
\begin{equation}
   {\cal E}(g)\sim\sum_{n=0}^\infty c_ng^n.
\label{three}
\end{equation}
The actual numerical values of~$c_n$s are given in~\cite{BDZ}
to~$n=8$ (to nine loops) from $D=0$ to~$D=4$ and we will use those
values.

In an analogy of the conventional field theory, we shall
call~${\cal E}(g)$ ``the vacuum energy density.'' The imaginary part
of the vacuum energy density is also evaluated in the leading order
bounce approximation~\cite{BDZ}:
\begin{equation}
   \Bigl[{\rm Im}\,{\cal E}(g)\Bigr]_{\rm bounce}=
   \frac{1}{(2\pi)^{D/2}}JM^{-1/2}
   \exp\left(-\frac{A_0}{g}\right),
\label{four}
\end{equation}
where
\begin{eqnarray}
   &&A_0=\frac{3}{2}\left(\frac{27\pi}{2}\right)^{D/2},\quad
   J=\left(\frac{3^D \pi^{D/2}}{g}\right)^{D/2},
\nonumber\\
   &&M=2\left[\left(\frac{2}{3}\right)^{D/2+1}\right]^D
   \exp\Sigma_2(D),\quad
   \Sigma_2(D)=\sum_{m=2}^\infty
   \frac{(m+D-1)!}{m!\,(D-1)!}\ln(1-3^{1-m}).
\label{bounce}
\end{eqnarray}
This formula tells us the leading large order behavior of the
perturbation series~(\ref{three})~\cite{BDZ}, or equivalently, the
position and nature of the Borel singularity nearest to the origin of
the Borel plane.

In completely the same reasoning as~\cite{SY}, we thus define the
Borel-Leroy transform of the perturbation series~$c_n$:
\begin{equation}
   B(z)=\sum_{n=0}^\infty\frac{c_n}{\Gamma(n+(D+1)/2)}z^n.
\label{five}
\end{equation}
The original vacuum energy density is supposed to be reproduced by
the Borel integral:
\begin{equation}
   {\cal E}(g)= \frac{1}{g^{(D+1)/2}}
	\int_0^\infty dz\,e^{-z/g} z^{(D+1)/2-1} B(z).
\label{six}
\end{equation}
The leading imaginary part~(\ref{four}) for~$g\ll1$ implies that
there exists a singularity of~$B(z)$ at~$z=A_0$. The argument of the
gamma function in~(\ref{five}) was chosen to sufficiently weaken this
singularity: The bounce singularity at~$z=A_0$ becomes a square root
branch point, irrespective of~$D$. The integration contour
of~(\ref{six}), therefore, must avoid the bounce singularity. This
deformation develops the imaginary part and we select the contour
along the upper side of the real axis.

The singularity at~$z=A_0$, however, limits the convergence radius of
the series~(\ref{five}). Consequently, in order to carry out the
integral~(\ref{six}) beyond~$z=A_0$, we have to execute the analytic
continuation of~$B(z)$. This continuation is possible only if we
completely know the Borel function~$B(z)$ within the convergence
circle (i.e., if all $c_n$s are known). However, we have only a
finite number of perturbation coefficients in actual.

To avoid this difficulty, we use the conformal mapping technique.
Introducing a new variable~$\lambda$ instead of~$z$,
\begin{eqnarray}
   z=4A_0\frac{\lambda}{(1+\lambda)^2},
\end{eqnarray}
new integration contour is confined in the convergence circle on
$\lambda$~plane. Indeed, the bounce singularity is transformed
to~$\lambda=1$ and new integration contour goes along inside the
upper arc of the convergence circle.

In this way, we have an imaginary part of the vacuum energy density
expressed by the perturbation coefficients~\cite{SY}:
\begin{eqnarray}
   &&\Bigl[{\rm Im}\,{\cal E}_0(g)\Bigr]_P
\nonumber\\
   &&=\left(\frac{A_0}{g}\right)^{(D+1)/2}
       \int_0^\pi d\theta\,
       \exp\left[-\frac{A_0}{g}\frac{1}{\cos^2(\theta/2)}\right]
       \frac{\sin(\theta/2)}{\cos^{D+2}(\theta/2)}
       \sum_{k=0}^Pd_k\sin(k\theta),
\label{formula}
\end{eqnarray}
with
\begin{equation}
   d_k=\sum_{n=0}^k(-1)^{k-n}
   \frac{\Gamma(k+n)(4A_0)^n}
        {(k-n)!\Gamma(2n)\Gamma(n+(D+1)/2)}c_n,
\label{seven}
\end{equation}
where $P$~denotes the order of approximation in this approach.

Now we present the result of~eq.~(\ref{formula}). When choosing the
presented range of the coupling constant~$g$ in Figures, we selected
as the upper bound the point that the effective mass changes it's
sign. To the one-loop order, we have
\begin{equation}
   \Sigma(p)\Bigr|_{p=0}
   =1-\frac{g}{2(4\pi)^{D/2}}+{\cal O}(g^2).
\end{equation}
Thus we choose as the upper bound of~$g$, $g=2(4\pi)^{D/2}$
(numerically, for $D=1$, $2$, $3$ and~$4$, it is $g=7.09$, $25.1$,
$89.1$ and~$315$, respectively). This upper bound might be regarded
as {\it an order of magnitude\/} estimation of the boundary of
``tunneling region''~\cite{SY}.

We plot eq.~(\ref{formula}) with $P=4$, $5$ and~$8$ as the function
of~$g$. The imaginary part is normalized by the result of the leading
bounce approximation~(\ref{four}). For~$D=0$~(Fig.~1), all the
perturbative coefficients in~(\ref{three}) can be analytically
computed because the partition function is just an ordinary integral.
Eq.~(\ref{formula}) rapidly converges to a uniform curve as shown
in~Fig.~1 and our prescription certainly seems to give the correct
imaginary part in~$D=0$.

For~$D=1$~(Fig.~2), three curves almost coincide over a wide range
of~$g$. In particular, the impressive agreement of $P=5$ and~$8$
indicates that the perturbation series to the fifth order is
practically sufficient to reproduce the correct imaginary part.
For~$D=2$~(Fig.~3), the range in which $P=5$ and~$8$ agree becomes
narrower and simultaneously the difference becomes larger. But still
eq.~(\ref{formula}) displays a rapid convergence for all~$g$ and
therefore $P=8$~seems to indicate the correct value. The order of
perturbation series required for a plausible evaluation of the
imaginary part is higher than the cases of $D=0$ and~$1$.

For $D=3$ and~$4$ (Figs.~4 and~5), there is a wild oscillation
in small~$g$ region. (For $g=0+$ the correct value is~$1$ because
the bounce result~(\ref{four}) becomes exact.) In spite of this
undesirable situation for the weak coupling, eq.~(\ref{formula})
seems converging in the strong coupling region. Therefore, in both
cases of~$D=3$ and~$4$, the correct imaginary part in the strong
coupling region seems to be smaller than the bounce result.

In summary, our method applied to the Gaussian propagator model seems
certainly working in lower dimensional cases, exhibiting a good
convergence. The results in turn indicate the correct imaginary part
of the vacuum energy density is much smaller than that of the leading
bounce approximation. On the other hand, the convergence in the
weak coupling region becomes worse in higher dimensional cases. We
guess this tendency holds in conventional field theories.

I would like to thank Dr.~H.~Suzuki for numerous discussions and
a careful reading of the manuscript. I am grateful to
Prof.~J.~Zinn-Justin for kind correspondence.

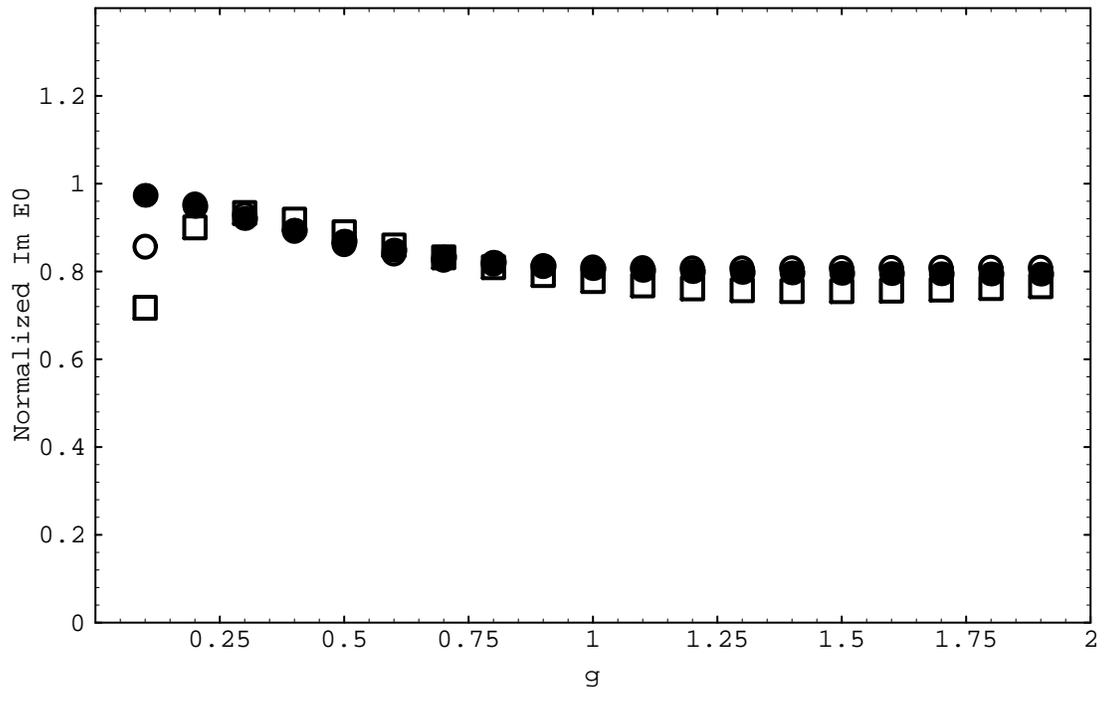
\begin{figure}
\caption{%
The imaginary part of the ``vacuum energy density'' evaluated by our
prescription~(\protect\ref{formula}) versus~$g$: $P=4$ (boxes),
$P=5$ (circles) and~$P=8$ (filled circles). The imaginary part is
normalized by the result of the leading bounce
approximation~(\protect\ref{four}).}
\label{fig1}
\end{figure}
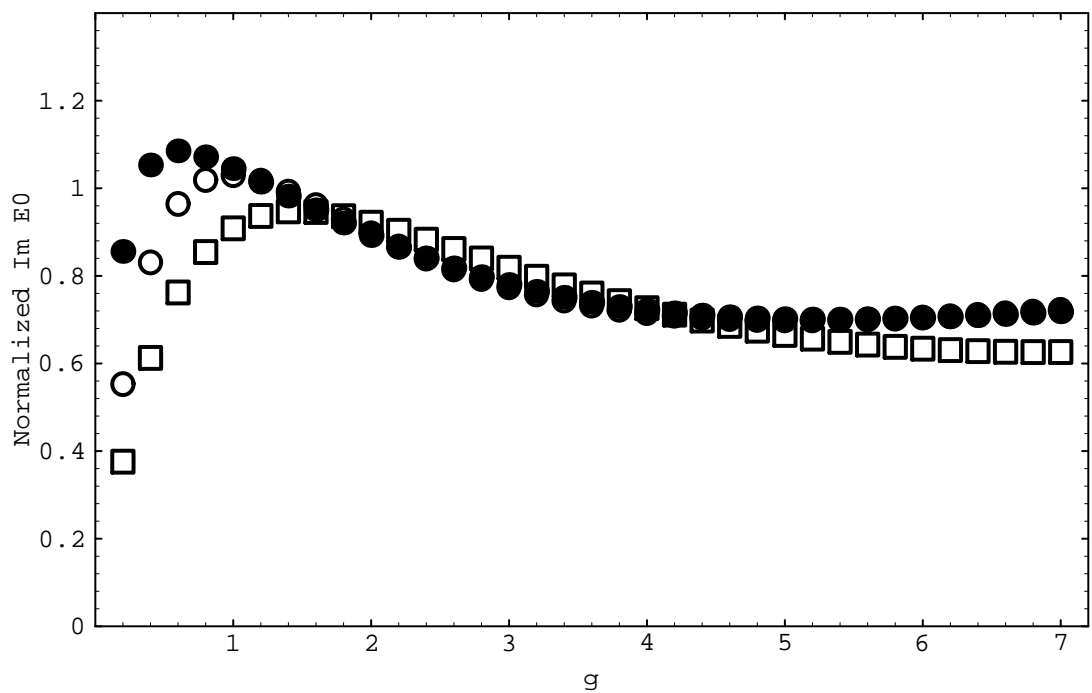
\begin{figure}
\caption{%
Same as Fig.~1 but for $D=1$.}
\label{fig2}
\end{figure}
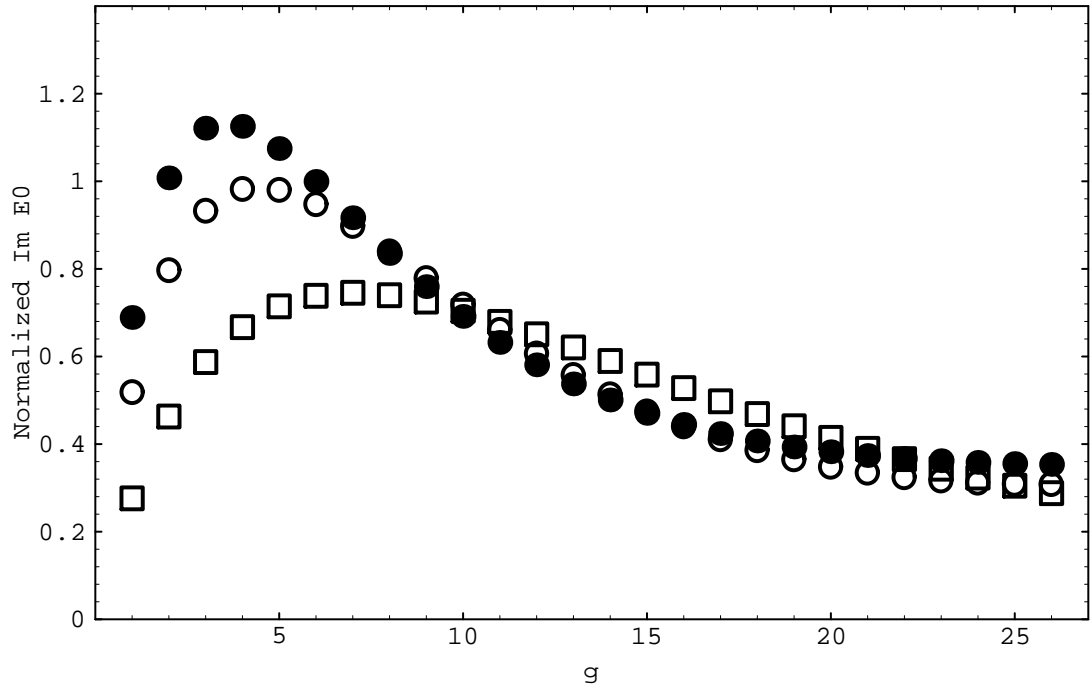
\begin{figure}
\caption{%
Same as Fig.~1 but for $D=2$.}
\label{fig3}
\end{figure}
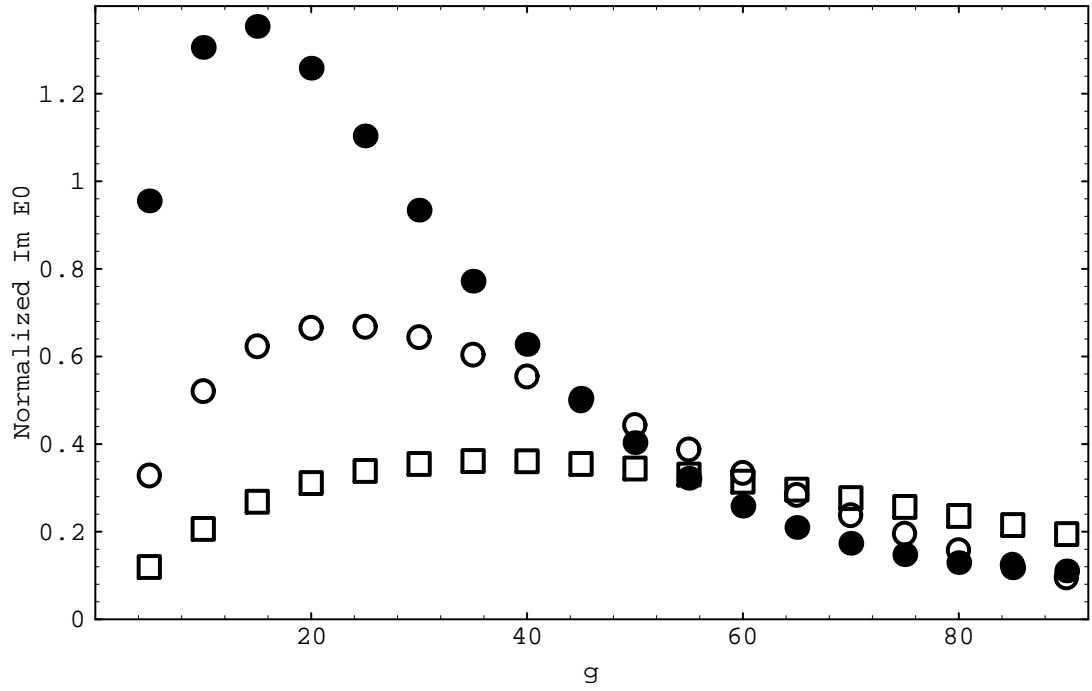
\begin{figure}
\caption{%
Same as Fig.~1 but for $D=3$.}
\label{fig4}
\end{figure}
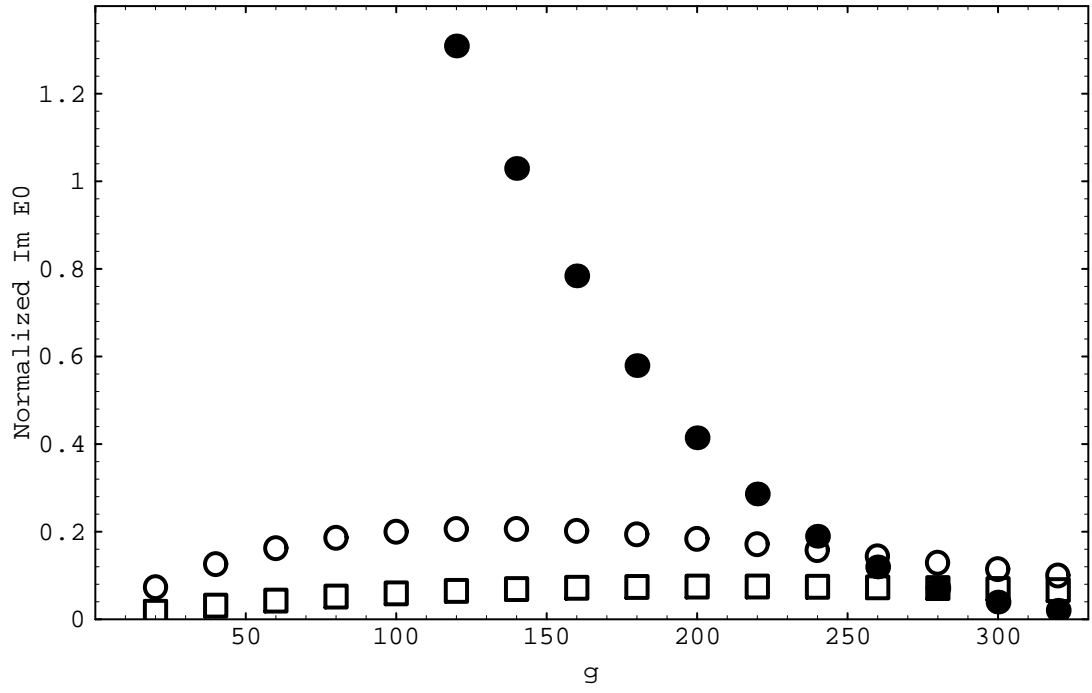
\begin{figure}
\caption{%
Same as Fig.~1 but for $D=4$.}
\label{fig5}
\end{figure}
\end{document}